\documentclass{iau}
\usepackage{graphicx} 

\title[IAUS291.~~SPAN512: A new mid-latitude pulsar survey] 
{SPAN512: A new mid-latitude pulsar survey with the Nan\c cay Radio Telescope} 

\author[G.~ Desvignes et al.]  
{Gregory Desvignes$^1$,
 Isma\"el Cognard$^2$,
 David Champion$^1$,\\
 Patrick Lazarus$^1$,
 Patrice Lespagnol$^3$,
 David A.~Smith$^4$\\
 \and Gilles Theureau$^3$
}

\affiliation{$^1$Max-Planck-Institut f\"ur Radioastronomie, Auf dem H\"ugel, 69 D-53121 Bonn, Germany\\ email: {\tt gdesvignes@mpifr-bonn.mpg.de} \\[\affilskip]
$^2$Laboratoire de Physique et Chimie de l'Environnement et de l'Espace, \\ 3A Avenue de la Recherche Scientifique, 45071 Orl\'eans cedex 2, France  \\[\affilskip]
$^3$Station de radioastronomie de Nan\c cay, Observatoire de Paris, CNRS/INSU, \\ 18330 Nan\c cay, France   \\[\affilskip]
$^4$Universit\'e Bordeaux 1, CNRS/IN2P3, CENBG Gradignan, 33175 Gradignan, France 
}

\pubyear{2012}
\volume{291}  
\jname{\mbox{Neutron Stars and Pulsars: Challenges and Opportunities after 80 years}}
\editors{J. van Leeuwen, ed.} 
\begin{document}

\maketitle

\begin{abstract}

We present an ongoing survey with the Nan\c cay Radio Telescope at L-band. The targeted
area is $74^\circ \lesssim  l <150^\circ$ and $3.5^\circ < |b| < 5^\circ$. 
This survey is characterized by a long integration
time (18 min), large bandwidth (512 MHz) and high time and frequency resolution (64 $\mu$s and
0.5 MHz) giving a nominal sensitivity limit of 0.055 mJy for long period pulsars. This is about 2
times better than the mid-latitude HTRU survey, and is designed to be complementary with
current large scale surveys. This survey will be more sensitive to transients (RRATs,
intermittent pulsars), distant and faint millisecond pulsars as well as scintillating sources (or
any other kind of radio faint sources) than all previous short-integration surveys. 

\end{abstract}


\firstsection 

\section{Introduction}
Major high-sensitivity pulsar surveys have recently started at different radio observatories due to 
improvement  of digital backends and computing resources over the past few years.
These L-Band surveys, ie. PALFA \cite{C06}, HTRU North \cite{B11} and South \cite{K10}, are concentrating their efforts at low galactic latitude ($|b|\lesssim 3.5^\circ$).
We present here a new survey  with the Nan\c cay Radio Telescope (NRT) at intermediate latitude outside the PALFA sky that started early 2012 and designed to be more sensitive than the HTRU mid-latitude survey.

\section{Observations}

Observations are made with the new state-of-the-art NUPPI backend
based on a CASPER\footnote{https://casper.berkeley.edu/} Roach
board. Compared to the previous BON instrument used for past surveys 
(e.g. \cite[Cognard \etal\ 2011 and Guillemot \etal\ 2012]{C11,G12}) the bandwidth is increased by a factor of 4. Also this versatile backend now performs a full Polyphase Filter Bank to mitigate the frequency leakage of RFIs.

The targeted field of view ($74^\circ \lesssim  l <150^\circ$ and $3.5^\circ < |b| < 5^\circ$ for $~230$ square degrees)  consists of a grid of $\sim 5800$ 18-min pointings recorded with a 64 $\mu$s time resolution with 1024 channels over the 512 MHz bandwidth.
Given the NRT system temperature and gain, we estimate the minimum flux density for long period pulsars to be 55-70 $\mu$Jy depending on the pointing declination. 
The basic parameters of the survey are listed in Table \ref{tab1}.






\begin{figure}
  \begin{center}
    \begin{minipage}{7.6cm}
     \begin{center}
      $\vcenter{\hbox{\includegraphics[height=7.5cm,angle=-90]{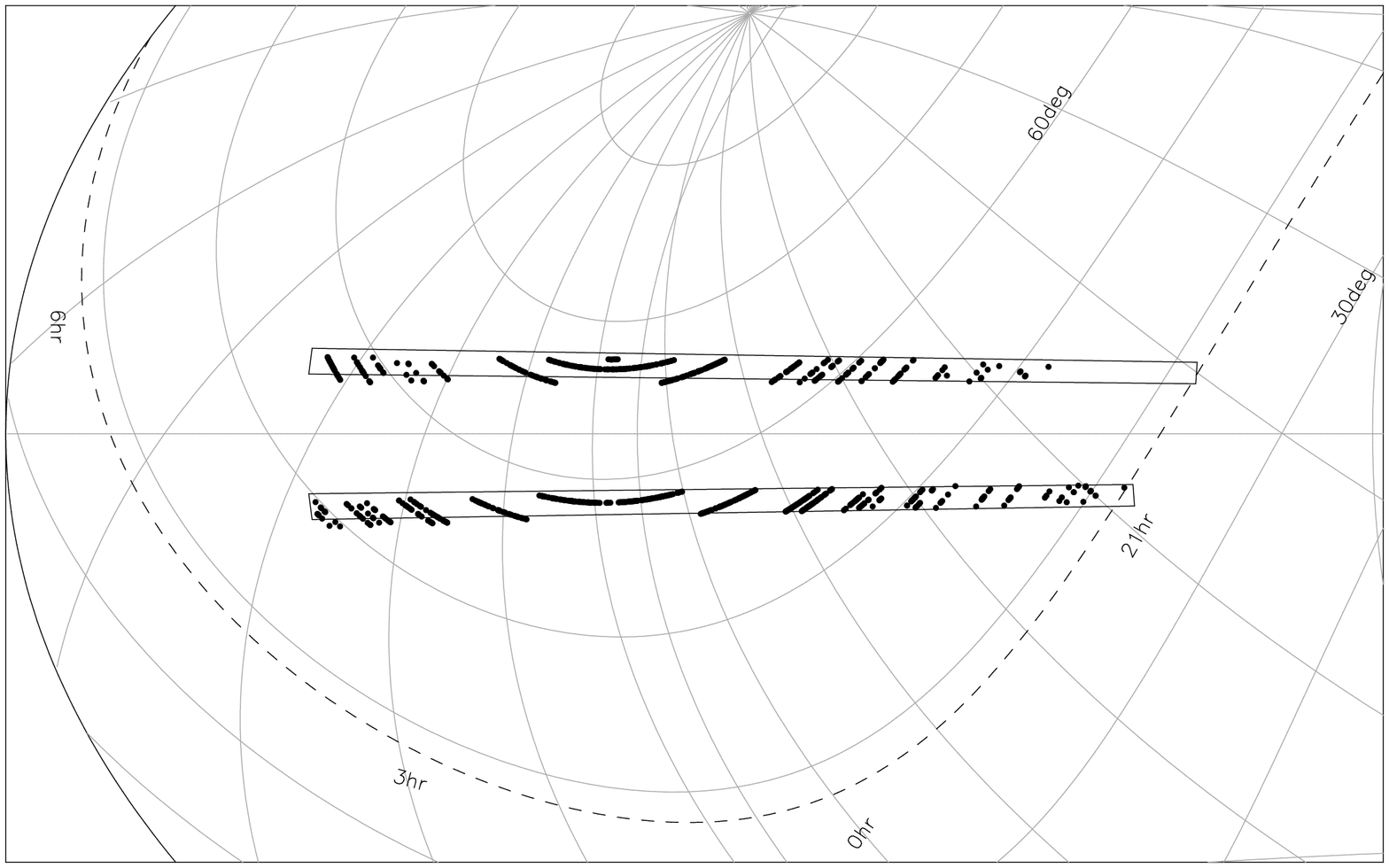}}}$
     \end{center}
    \end{minipage}
    \begin{minipage}{5.9cm}
     \begin{center}
      $\vcenter{\hbox{\includegraphics[trim=30 0 0 0,
             clip, height=5.7cm]{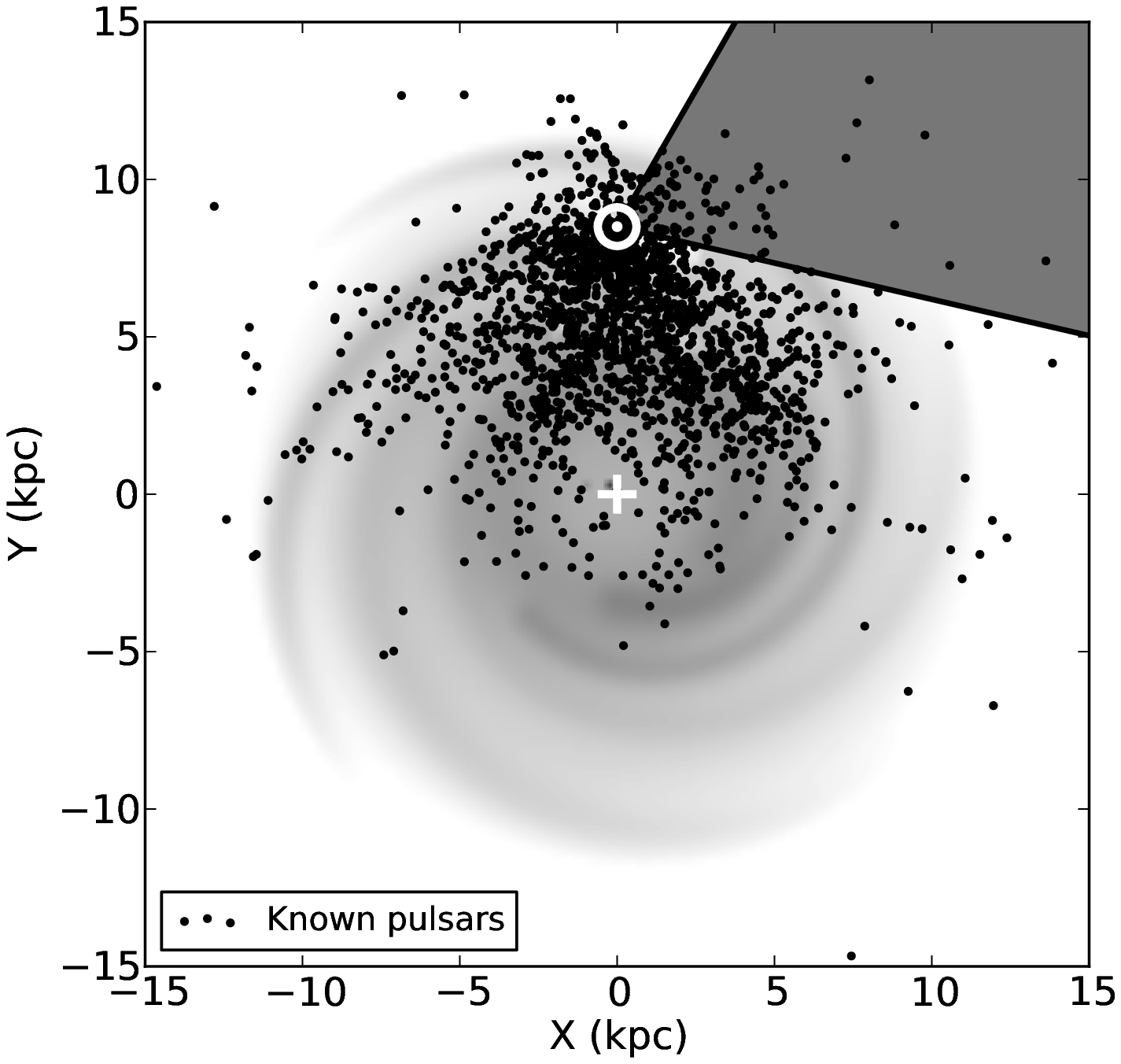}}}$
     \end{center}
    \end{minipage}
    \caption{{\it Left panel:} View of the Galactic Plane in galactic coordinates.
    The two black boxes delimit the SPAN512 area and the black dots
    represent the 859  pointings made to date. The dashed line shows the Arecibo North declination limit. 
    {\it Right panel:} View of the Galactic Plane from the North
    Galactic Pole. The Galactic Center is located at (0,0).
    The gray scale is computed from the NE 2001 electron
    density model by \cite{C02}.}
    \label{fig1}
  \end{center}
\end{figure}

\begin{table}
  \begin{center}
    \caption{Parameters of the SPAN survey}
    \label{tab1}
    \begin{tabular}{ll}
    \hline
    Sampling time		& 64 $\mu$s \\
    Total bandwidth		& 512 MHz \\
    Number of channels		& 1024 \\
    Center frequency		& 1486 MHz \\
    Integration time		& 18 min \\
    Final quantization		& 4 bits \\
    Gain			& 1.4 K/Jy \\
    System temperature		& 35 K \\
    Nominal sensitivity		& 0.055 mJy \\
    Total observing time~~~~~~	& 1740 hours \\
    \hline

    \end{tabular}
  \end{center}
\end{table}

\section{Processing}
A total data volume of 50 TB is expected after completion of
this program. To search these data, 2 different schemes are
considered, both using the Presto\footnote{https://github.com/scottransom/presto}
package:

\begin{itemize}
\item A quicklook pipeline that reduces the original time resolution
by a factor of 4 is used on site and is able to keep up with the
data acquisition. No acceleration search is performed at this
stage.

\item To search the full-resolution data with acceleration, we are currently
implementing a new pipeline\footnote{https://github.com/plazar/pipeline2.0}
to run at the IN2P3 Computing Center\footnote{http://cc.in2p3.fr/}
developed originally for the ongoing PALFA survey. To
remain sensitive to very short orbital period binaries, we also  split the observations in half and analyze each separately before combining results.
In the light of the recent discoveries of highly dispersed radio bursts (\cite[Lorimer \etal\ 2007 and
Keane \etal\ 2012]{L07,K12}), the data are searched for single pulses up to a DM of 1800 cm$^{-3}$ pc. 
\end{itemize}

\section{Conclusion}

Preliminary results
of the quicklook pipeline indicate a relatively low RFI
environment ($\sim $10\% of the radio band being masked),
especially considering the very wide bandwidth. 
In the observations to date, three previously known pulsars (including one MSP) were redetected.

About 860 of the 5800 planned pointings  have been made (15\% of the $~230$ square degrees of this survey) and completion is 
expected by 2013 with the discovery of 10 to 30 new sources according to population models \cite{L06}.

\acknowledgments
The Nan\c cay Radio Telescope is part of the Paris Observatory, associated with the Centre National de la
Recherche Scientifique (CNRS) and the University of Orl\'eans, France.
We also acknowledge the Centre de Calcul CC-IN2P3. (IN2P3, CNRS Villeurbanne, France) for providing us
with computing resources.

\end{document}